\documentclass[12pt]{article}

\usepackage{amsmath,amssymb,amsfonts,amsthm}

\begin{document}

\begin{flushright} Statistical Inference for Stochastic Processes, \\
25, no. , pp. , 2022, \  DOI: 10.1007/s11203-022-09278-4.
\end{flushright}

\vskip 0.8cm

\begin{center} { M. V. BURNASHEV
\footnote[1]{Burnashev M. V. is with Institute for Information Transmission
Problems, Russian Academy of Sciences, Moscow; email: burn@iitp.ru} \\
{ON STEIN'S LEMMA IN HYPOTHESES TESTING IN GENERAL NON-ASYMPTOTIC CASE}}
\footnote[2]{This work was supported  by the Russian Foundation
for Basic Research under Grant 19-01-00364.}
\end{center}

{\bf Abstract -- }
The problem of testing two simple hypotheses in a general probability
space is considered. For a fixed type-I error probability, the best
exponential decay rate of the type-II error probability is
investigated. In regular asymptotic cases (i.e., when the length
of the observation interval grows without limit) the best decay rate
is given by Stein's exponent. In the paper, for a general probability
space, some non-asymptotic lower and upper bounds for the best rate
are derived. These bounds represent pure analytic relations without
any limiting operations. In some natural cases, these bounds also
give the convergence rate for Stein's exponent. Some illustrating
examples are also provided.


\section{Introduction and Main result}
We assume that there are given probability measures $\mathbf P$
and $\mathbf Q$ on a measurable space $(\cal X,{\cal B})$.
We consider the problem of testing the simple hypothesis
$\mathcal H_{0}$ against the simple alternative
$\mathcal H_{1}$, based on observation
${\mathbf x}$:
\begin{equation}\label{mod2}
\begin{gathered}
\mathcal H_{0}: {\mathbf x} \sim {\mathbf P}, \\
\mathcal H_{1}: {\mathbf x} \sim {\mathbf Q}.
\end{gathered}
\end{equation}

If for the testing of hypotheses $\mathcal H_{0}$ and $\mathcal H_{1}$
a decision (acceptance) region ${\mathcal D} \in {\cal X}$ is
chosen, such that
\begin{equation}\label{testD1}
\begin{gathered}
{\mathbf x} \in \mathcal D \Rightarrow \mathcal H_{0}, \qquad
{\mathbf x} \not\in \mathcal D \Rightarrow \mathcal H_{1},
\end{gathered}
\end{equation}
then the type-I error probability $\alpha({\mathcal D})$ and the
type-II error probability $\beta({\mathcal D})$ are defined by
formulas, respectively,
\begin{equation}\label{defalpha21}
\alpha({\mathcal D}) =
\mathbf{P}({\mathbf x} \not\in \mathcal D|\mathcal H_{0})
\end{equation}
and
\begin{equation}\label{defbeta21}
\begin{gathered}
\beta({\mathcal D}) =
\mathbf{P}({\mathbf x} \in \mathcal D|\mathcal H_{1}).
\end{gathered}
\end{equation}

In the paper, we consider the case when the type-I error probability $\alpha$,
$0 < \alpha < 1$, is fixed, and we are interested in the minimal possible type-II
error probability $\beta(\alpha)$
\begin{equation}\label{defbeta3a}
\beta(\alpha) =
\inf\limits_{{\mathcal D}:\alpha({\mathcal D}) \leq \alpha}
\beta({\mathcal D}).
\end{equation}
The corresponding optimal decision region ${\mathcal D}(\alpha)$
to this problem is given by Neyman-Pearson lemma \cite{Wald, Lehmann}.

We will need the following notion \cite{Kullback}.

{\bf Definition 1}. \textit{
For probability distributions ${\mathbf P}$ and ${\mathbf Q}$ on a
measurable space $(\cal X,{\cal B})$ the function $D({\mathbf P}||{\mathbf Q})$
(Kullback--Leibler distance or divergence for measures ${\mathbf P}$
and ${\mathbf Q}$) is defined as
\begin{equation}\label{Stein0}
\begin{gathered}
D({\mathbf P}||{\mathbf Q}) = {\mathbf E}_{\mathbf P}
\ln\frac{d{\mathbf P}}{d{\mathbf Q}}({\mathbf x}) \geq 0,
\end{gathered}
\end{equation}
where the expectation is taken over the measure ${\mathbf P}$.}

In order to describe Stein's lemma and its relation to the value
$\beta(\alpha)$ via the function $D({\mathbf P}||{\mathbf Q})$, we consider
first the case ${\cal X} = {\mathbf R}^{n}$, when an observation ${\mathbf x}$
from \eqref{mod2} has the form ${\mathbf x} = {\mathbf x}_{n} =
(x_{1},\ldots,x_{n}) \in {\mathbf R}^{n}$. We assume that the sample
${\mathbf x}_{n}$ consists of independent and identically distributed (i.i.d.)
random variables $x_{i}$, $i=1,\ldots,n$ with a given distribution
${\mathbf P}$ (in the case of $\mathcal H_{0}$), or with a given distribution
${\mathbf Q}$ (in the case of $\mathcal H_{1}$):
\begin{equation}\label{mod1}
\begin{gathered}
\mathcal H_{0}: x_{i} \sim {\mathbf P}, \qquad i=1,\ldots,n; \\
\mathcal H_{1}: x_{i} \sim {\mathbf Q}, \qquad i=1,\ldots,n.
\end{gathered}
\end{equation}

If, in model \eqref{mod1} ${\mathbf P} \not\equiv {\mathbf Q}$,
then $D({\mathbf P}||{\mathbf Q}) > 0$ and the value $\beta(\alpha)$
decreases exponentially as $n \to \infty$. Moreover, we have
\cite[Theorem 3.3]{Kullback}
\begin{equation}\label{Stein01}
\begin{gathered}
\lim_{\alpha\to 0}\lim_{n\to \infty}\frac{1}{n}\ln\beta(\alpha) =
-D({\mathbf P}||{\mathbf Q}).
\end{gathered}
\end{equation}

Relation \eqref{Stein01} is called Stein's lemma \cite{Stein},
\cite{Chernov56}.
\footnote{The reference \cite{Stein} is due to remarks from
\cite[p. 18]{Chernov56} and  \cite[Theorem 3.3]{Kullback} with attribute to
the unpublished paper of C. Stein.}
In the case of i.i.d. random variables, its proof can be found in
\cite[Theorem 3.3]{Kullback}, \cite[Theorem 12.8.1]{CT}. It is natural to expect
that the relation \eqref{Stein01} holds not only for i.i.d. random variables
as in model \eqref{mod1}, but in much more general cases.
Following \cite[Theorem 12.8.1]{CT}, some particular
analogs of formula \eqref{Stein01} have already appeared for the cases of
stationary Gaussian \cite{ZhangPoor11} and Poisson \cite{Bur21} random processes
(with observation time $T$ instead of $n$).

Note that the relation \eqref{Stein01} is essentially oriented to the case when
the type-I and the type-II errors imply very different losses for us, and we are
mainly interested in minimization of the type-II error probability
$\beta = {\mathbf P}\{H_{0}|H_{1}\}$. The case is quite popular in many
applications (see, e.g., \cite{ZhangPoor11} and references therein). Also,
Chernov \cite[p. 17]{Chernov56} wrote on such case: ``It occasionally happens in
practice that it is important to obtain $\beta$ very small, whereas a relatively
large value of $\alpha$, like $.05$ or $.10$, is not disastrous''

In this paper, for the general model \eqref{mod2} some non-asymptotic
generalization and strengthening of the relation \eqref{Stein01} is derived.
Such non-asymptotic results allowed to generalize the relation \eqref{Stein01}
for testing a simple hypothesis versus a composite one \cite{Bur21}. Also,
such non-asymptotic results would allow essentially simplify similar results
for Gaussian processes \cite{ZhangPoor11, Bur17a}.

\subsection{Assumption}
In order to simplify formulation of the results and to avoid pathological
cases, we assume that the following assumption is satisfied:

{\bf I}. Probability measures $\mathbf P$
and $\mathbf Q$ on a measurable space $(\cal X,{\cal B})$ are equivalent and
Radon-Nikodim derivative $d{\mathbf P}/d{\mathbf Q}({\mathbf x})$ for them is
defined, positive and finite for ${\mathbf x} \in {\cal X}$.

\subsection{Main result}

The main result of this paper is the following.

{\bf Theorem}. \textit{If the assumption ${\bf I}$ holds, then
the minimal possible $\beta(\alpha)$,
$0 < \alpha<1$, satisfies the bounds
%(see \eqref{Stein0})
\begin{equation}\label{Stein11}
\begin{gathered}
\ln\beta(\alpha)\geq -\frac{D({\mathbf P}||{\mathbf Q})+ h(\alpha)}{1-\alpha},\\
h(\alpha) = -\alpha\ln\alpha - (1-\alpha)\ln(1-\alpha),
\end{gathered}
\end{equation}
and
\begin{equation}\label{Stein12}
\begin{gathered}
\ln\beta(\alpha) \leq -D({\mathbf P}||{\mathbf Q}) + \mu_{0}(\alpha),
\end{gathered}
\end{equation}
where $\mu_{0}(\alpha)$ is the minimal value $\mu_{0}$, satisfying the relation}
\begin{equation}\label{defmu0}
\begin{gathered}
{\mathbf P}_{{\mathbf P}}\left\{
\ln\frac{d{\mathbf P}}{d{\mathbf Q}}({\mathbf x}) \leq
D({\mathbf P}||{\mathbf Q})-\mu_{0}\right\} \leq \alpha.
\end{gathered}
\end{equation}

With some loss of accuracy, we can simplify the upper bound
\eqref{Stein12}, using some upper bound for the value
$\mu_{0}(\alpha)$ (see \eqref{Cheb1} below). For example, we get

{\bf Corollary}. \textit{The minimal possible $\beta(\alpha)$,
$0 < \alpha<1$, satisfies also the upper bound
\begin{equation}\label{Stein14}
\begin{gathered}
\ln\beta(\alpha) \leq -D({\mathbf P}||{\mathbf Q}) + \alpha^{-1/2}
r_{1}({\mathbf P},{\mathbf Q}),
\end{gathered}
\end{equation}
where}
\begin{equation}\label{rem1}
\begin{gathered}
r_{1}({\mathbf P},{\mathbf Q}) = \left\{{\mathbf E}_{{\mathbf P}}\left[
\ln\frac{d{\mathbf P}}{d{\mathbf Q}}({\mathbf x})\right]^{2} -
D^{2}({\mathbf P}||{\mathbf Q})
\right\}^{1/2}.
\end{gathered}
\end{equation}

Note that all bounds \eqref{Stein11}, \eqref{Stein12} and
\eqref{Stein14} are pure analytical relations without any limiting
operations. Also, for a fixed $\alpha$, both lower bound \eqref{Stein11}
and upper bound \eqref{Stein14} are close to each other, if the value
$r_{1}({\mathbf P},{\mathbf Q})$ is much smaller than
$D({\mathbf P}||{\mathbf Q})$.

{\it Remark 1}. For the particular case of Poisson processes Theorem was
proved in \cite{Bur21} (as Theorem 3).

{\it Remark 2}. Note that if in some asymptotic setting the Central Limit
Theorem is applicable to the random variable
$D({\mathbf P}||{\mathbf Q})-\ln(d{\mathbf P}/d{\mathbf Q})({\mathbf x})$,
then the remaining term $r_{1}({\mathbf P},{\mathbf Q})$ in \eqref{Stein14}
has the optimal order (in $D({\mathbf P}||{\mathbf Q})$) for fixed $\alpha$ and
large $D({\mathbf P}||{\mathbf Q})$ (see Example 3 below).

\section{Proofs}

\subsection{Proof of Theorem }
Derive first the lower bound \eqref{Stein11}. Without loss of generality and
in order to avoid bulky formulas, note that using a natural randomization on
the ``bound'' of a decision region ${\mathcal D} \in {\cal X}$,
we have with ${\cal D}^{c} = {\cal X} \setminus {\cal D}$ and
$\beta=\beta({\cal D})$, $\alpha = \alpha({\cal D})$
\begin{equation}\label{Defab}
\begin{gathered}
\beta = {\mathbf Q}({\cal D}) = \int\limits_{{\cal D}}
\frac{d{\mathbf Q}}{d{\mathbf P}}({\mathbf x})
d{\mathbf P}({\mathbf x}),  \qquad
\alpha = {\mathbf P}({\cal D}^{c}).
\end{gathered}
\end{equation}
Since ${\mathbf P}({\cal D})=1-{\mathbf P}({\cal D}^{c}) = 1-\alpha$,
then considering $d{\mathbf P}/(1-\alpha)$
as the probability distribution on ${\cal D}$, and using the inequality
$\ln {\mathbf E}\xi \geq {\mathbf E}\ln \xi$, we have
\begin{equation}\label{Stein1a}
\begin{gathered}
\ln\frac{\beta}{1-\alpha} = \ln\left[\frac{1}{(1-\alpha)}
\int\limits_{{\cal D}}
\frac{d{\mathbf Q}}{d{\mathbf P}}({\mathbf x})
d{\mathbf P}({\mathbf x})\right] \geq \\
\geq \frac{1}{(1-\alpha)}\int\limits_{{\cal D}}
\ln\frac{d{\mathbf Q}}{d{\mathbf P}}({\mathbf x})
d{\mathbf P}({\mathbf x}) = \\
= - \frac{D({\mathbf P}||{\mathbf Q})}{1-\alpha} -
\frac{1}{(1-\alpha)}\int\limits_{{\cal D}^{c}}
\ln\frac{d{\mathbf Q}}{d{\mathbf P}}({\mathbf x})
d{\mathbf P}({\mathbf x}).
\end{gathered}
\end{equation}
Since ${\mathbf P}({\cal D}^{c})=\alpha$, similarly to \eqref{Stein1a},
the last term in the right-hand side of \eqref{Stein1a} gives
\begin{equation}\label{Stein1b}
\begin{gathered}
\frac{1}{(1-\alpha)}\int\limits_{{\cal D}^{c}}
\ln\frac{d{\mathbf Q}}{d{\mathbf P}}({\mathbf x})
d{\mathbf P}({\mathbf x}) \leq \\
\leq \frac{\alpha}{(1-\alpha)}\ln\left[\frac{1}{\alpha}
\int\limits_{{\cal D}^{c}}d{\mathbf Q}({\mathbf x})\right] =
\frac{\alpha}{(1-\alpha)}\ln\frac{1-\beta}{\alpha} \leq
\frac{\alpha}{(1-\alpha)}\ln\frac{1}{\alpha}.
\end{gathered}
\end{equation}
Therefore, from \eqref{Stein1a} and \eqref{Stein1b} we have
\begin{equation}\label{Stein1c}
\begin{gathered}
\ln\frac{\beta}{1-\alpha} \geq
- \frac{D({\mathbf P}||{\mathbf Q})}{1-\alpha} -
\frac{\alpha}{(1-\alpha)}\ln\frac{1}{\alpha},
\end{gathered}
\end{equation}
from where the lower bound \eqref{Stein11} follows.

In order to prove the upper bound \eqref{Stein12}, we set a value
$\mu > 0$, and define the acceptance region in favor of ${\mathbf P}$
\begin{equation}\label{SteinP1}
\begin{gathered}
{\cal A}_{\mu}= \left\{{\mathbf x} \in {\cal X}:
\ln\frac{d{\mathbf P}}{d{\mathbf Q}}({\mathbf x})
\geq D({\mathbf P}||{\mathbf Q})-\mu\right\}.
\end{gathered}
\end{equation}
Denote by $\alpha_{\mu}$ and $\beta_{\mu}$ the type-I and
the type-II error probabilities for the acceptance region
${\cal A}_{\mu}$, respectively. Then by \eqref{SteinP1}
\begin{equation}\label{SteinQ1}
\begin{gathered}
\beta_{\mu} = \int\limits_{{\cal A}_{\mu}}d{\mathbf Q} =
\int\limits_{{\cal A}_{\mu}}
\frac{d{\mathbf Q}}{d{\mathbf P}}({\mathbf x})d{\mathbf P} \leq
e^{-D({\mathbf P}||{\mathbf Q})+\mu}.
\end{gathered}
\end{equation}
Also
\begin{equation}\label{SteinP1a}
\begin{gathered}
\alpha_{\mu} = {\mathbf P}_{{\mathbf P}}\left\{
\ln\frac{d{\mathbf P}}{d{\mathbf Q}}({\mathbf x}) \leq
D({\mathbf P}||{\mathbf Q})-\mu\right\} ={\mathbf P}(\eta \geq \mu),
\end{gathered}
\end{equation}
where
\begin{equation}\label{defeta}
\eta = D({\mathbf P}||{\mathbf Q})-
\ln\frac{d{\mathbf P}}{d{\mathbf Q}}({\mathbf x}), \qquad
{\mathbf E}\eta =0.
\end{equation}
The best solution is to set $\mu$ such that $\alpha_{\mu}=\alpha$.
Therefore, we define $\mu_{0}=\mu_{0}(\alpha)$ by the formula
\eqref{defmu0}. Then we have
\begin{equation}\label{SteinQ3b}
\begin{gathered}
\alpha_{\mu_{0}(\alpha)} = \alpha,
\end{gathered}
\end{equation}
and by \eqref{SteinQ1}
\begin{equation}\label{SteinQ2b}
\begin{gathered}
\ln\beta_{\mu_{0}(\alpha)} \leq -D({\mathbf P}||{\mathbf Q}) +
\mu_{0}(\alpha),
\end{gathered}
\end{equation}
which proves the upper bound \eqref{Stein12}.

\subsection{Proof of Corollary}
In order to prove the upper bound \eqref{Stein14}, we simplify the
relation \eqref{defmu0}, using the classic Chebyshev inequality
\begin{equation}\label{Cheb1}
\begin{gathered}
{\mathbf P}(\eta \geq \mu) \leq \mu^{-2}{\mathbf E}(\eta_{+}^{2}),
\end{gathered}
\end{equation}
where $z_{+} = \max\{0,z\}$. Then from \eqref{SteinP1a} and \eqref{Cheb1}
\begin{equation}\label{Cheb1a}
\begin{gathered}
\alpha_{\mu} \leq \mu^{-2}{\mathbf E}_{\mathbf P}\left[
D({\mathbf P}||{\mathbf Q})-
\ln\frac{d{\mathbf P}}{d{\mathbf Q}}({\mathbf x})\right]_{+}^{2}.
\end{gathered}
\end{equation}
We set $\mu = \mu_{1}> 0$ as follows (see \eqref{rem1})
\begin{equation}\label{Cheb2}
\begin{gathered}
\mu_{1} = \alpha^{-1/2}\left\{{\mathbf E}_{\mathbf P}\left[
D({\mathbf P}||{\mathbf Q})-
\ln\frac{d{\mathbf P}}{d{\mathbf Q}}({\mathbf x})\right]_{+}^{2}
\right\}^{1/2} = \\
= \alpha^{-1/2}r_{1}({\mathbf P},{\mathbf Q}).
\end{gathered}
\end{equation}
Then we get
\begin{equation}\label{SteinQ3}
\begin{gathered}
\alpha_{\mu_{1}} \leq \alpha,
\end{gathered}
\end{equation}
and by \eqref{SteinQ1} and \eqref{Cheb2}
\begin{equation}\label{SteinQ2}
\begin{gathered}
\ln\beta_{\mu_{1}} \leq -D({\mathbf P}||{\mathbf Q})+\mu_{1} = \\
= -D({\mathbf P}||{\mathbf Q}) +
\alpha^{-1/2}r_{1}({\mathbf P},{\mathbf Q}),
\end{gathered}
\end{equation}
which proves the upper bound \eqref{Stein14}. \qquad $\Box$

\section{Examples}

{\bf Example 1 (Poisson processes)}.
Denote by ${\mathbf x}_{T}=x_{{\mathbf u}_T} =
\{x_{{\mathbf u}_T}(t), 0 \leq t \leq T\}$, the Poisson process with the
intensity ${\mathbf u}_T = \{u_{T}(t), 0 \leq t \leq T\}$ Kutoyants (1984)
and let ${\cal X}_{T}$ denote the observation space (i.e.
${\mathbf x}_{T} \in {\cal X}_{T}$). Consider testing a simple hypothesis
$\mathcal H_{0}$ versus a simple alternative $\mathcal H_{1}$, based on
observations ${\mathbf x}_T = \{{\mathbf x}_{T}(t), 0 \leq t \leq T\}$:
\begin{equation}\label{mod1a}
\begin{split}
&\mathcal H_{0}: {\mathbf x}_{T} = x_{{\mathbf p}_T}, \\
&\mathcal H_{1}: {\mathbf x}_{T} = x_{{\mathbf q}_T},
\end{split}
\end{equation}
where intensities ${\mathbf p}_T$ and ${\mathbf q}_T$ are given.

Poisson processes on $[0,T]$ with intensities ${\mathbf p}_{T}$ and
${\mathbf q}_{T}$ generate on $({\cal X}_{T},{\cal B}_{T})$ measures
${\mathbf P}_{{\mathbf p}_{T}}$ and ${\mathbf P}_{{\mathbf q}_{T}}$, respectively.
The following formula for the Radon--Nikodim derivative holds
(see, e.g., Kutoyants (1984))
\begin{equation}\label{radnic}
\begin{gathered}
\frac{d{\mathbf P}_{{\mathbf p}_{T}}}{d{\mathbf P}_{{\mathbf q}_{T}}}
({\mathbf x}) = \exp\left\{(\ln ({\mathbf p}_{T}/{\mathbf q}_{T}),dx) -
({\mathbf p}_{T}-{\mathbf q}_{T},{\mathbf 1})\right\},
\end{gathered}
\end{equation}
where for functions ${\mathbf f} = \{f(t), 0 \leq t \leq T\}$,
${\mathbf s} = \{s(t), 0 \leq t \leq T\}$ and Poisson process
$x(t)$ we used the notation
\begin{equation}\label{not1}
({\mathbf f},{\mathbf s}) = \int\limits_{0}^{T}f(t)s(t)dt, \qquad
({\mathbf f}, dx) = \int\limits_{0}^{T}f(t)dx(t).
\end{equation}
It is supposed that the right-hand side of \eqref{radnic} is well defined and
therefore the measures ${\mathbf P}_{{\mathbf p}_{T}}$ and
${\mathbf P}_{{\mathbf q}_{T}}$ are equivalent (see, e.g., Kutoyants (1984)).

From (\ref{radnic}) for any integrable function
${\mathbf f}_{T}=\{f_{T}(t), 0 \leq t \leq T\}$
the "expectation" formula follows (see, e.g., Kutoyants (1984))
\begin{equation}\label{radnic2}
\ln {\mathbf E}_{{\mathbf p}_{T}}e^{(\ln{\mathbf f}_{T},dx)} =
({\mathbf f}_{T}-{\mathbf 1},{\mathbf p}_{T}),
\end{equation}
where ${\mathbf E}_{{\mathbf p}_{T}}$ means expectation with
respect to the measure ${\mathbf P}_{{\mathbf p}_{T}}$.
In particular, from \eqref{radnic2} we get
\begin{equation}\label{radnic3}
\begin{gathered}
{\mathbf E}_{{\mathbf p}_{T}}({\mathbf f}_{T},dx) =
({\mathbf f}_{T},{\mathbf p}_{T}), \\
{\mathbf E}_{{\mathbf p}_{T}}({\mathbf f}_{T},dx)^{2} =
({\mathbf f}_{T},{\mathbf p}_{T})^{2}+ ({\mathbf f}_{T}^{2},{\mathbf p}_{T}).
\end{gathered}
\end{equation}
Then, from \eqref{Stein0}, \eqref{radnic} and \eqref{radnic3} we have
\begin{equation}\label{Stein1a00}
\begin{gathered}
D({\mathbf P}_{{\mathbf p}_{T}}||{\mathbf P}_{{\mathbf q}_{T}}) =
D({\mathbf p}_{T}||{\mathbf q}_{T}) =
(\ln({\mathbf p}_{T}/{\mathbf q}_{T}),{\mathbf p}_{T}) -
({\mathbf p}_{T}-{\mathbf q}_{T},{\mathbf 1}).
\end{gathered}
\end{equation}
For the remaining term $r_{1}({\mathbf P},{\mathbf Q})$ from
\eqref{Stein14}-\eqref{rem1} we get by \eqref{radnic3} and \eqref{Stein1a00}
\begin{equation}\label{remex1}
\begin{gathered}
r_{1}({\mathbf P},{\mathbf Q}) = \left\{
(\ln^{2}({\mathbf p}_{T}/{\mathbf q}_{T}),{\mathbf p}_{T})\right\}^{1/2},
\end{gathered}
\end{equation}
which gives an estimate to the convergence rate in \eqref{Stein01}
(provided the right-hand side of \eqref{remex1} is finite).

{\bf Example 2 (Gaussian processes)}.
Note that by the Karhunen-Lo\'{e}ve expansion most of detection problems on
continuous-time Gaussian processes can be reduced to the discrete-time case
with Gaussian random vectors \cite[Ch. VI]{Poor}. Formally, such representation
consists of infinite number of expansion terms, but in practice such infinite
series can be well approximated by a finite summation (without essential loss
for performance characteristics).

For that reason we consider
testing of simple hypotheses $\mathcal H_{0}$ and $\mathcal H_{1}$, based on
observation of Gaussian random vectors ${\mathbf x}_{n} \in {\mathbf R}^{n}$:
\begin{equation}\label{mod1aa}
\begin{gathered}
\mathcal H_{0}: {\mathbf x}_{n} = \boldsymbol{\xi}_{n}, \qquad
\boldsymbol{\xi}_{n} \sim
{\mathcal N}(\boldsymbol{0},\mathbf{I}_{n}), \\
\mathcal H_{1}: {\mathbf x}_{n} = \boldsymbol{\eta}_{n}, \qquad
\boldsymbol{\eta}_{n} \sim
{\mathcal N}(\boldsymbol{0},\mathbf{M}_{n}),
\end{gathered}
\end{equation}
where the sample $\boldsymbol{\xi}_{n} = (\xi_{1},\ldots,\xi_{n})$
represents ``noise'' and consists of independent and identically
distributed Gaussian random variables with zero means and variances $1$. The
stochastic ``signal'' $\boldsymbol{\eta}_{n} = (\eta_{1},\ldots,\eta_{n})$ is a
Gaussian random vector with zero mean and covariance matrix
$\mathbf{M}_{n}$ (which is symmetric and nonnegative-definite).

Denote by ${\mathbf P}_{n}$ and ${\mathbf Q}_{n}$ distributions of
$\boldsymbol{\xi}_{n}$ and $\boldsymbol{\eta}_{n}$, respectively.
Note that if $\det\mathbf{M}_{n} = 0$, then distributions
${\mathbf P}_{n}$ and ${\mathbf Q}_{n}$ are orthogonal, and therefore,
hypotheses $\mathcal H_{0}$ and $\mathcal H_{1}$ can be tested with
$\alpha = \beta =0$.

Hence, we assume that $\det\mathbf{M}_{n} > 0$, and then
$\mathbf{M}_{n}$ is a symmetric and positive-definite matrix.
Denote by $\lambda_{1},\ldots,\lambda_{n}$ the eigenvalues of
$\mathbf{M}_{n}$ (all positive).

In order to simplify calculations, we reduce model \eqref{mod1aa}
to the equivalent one with the diagonal matrix $\mathbf{M}_{n}$.
Indeed, for the covariance matrix $\mathbf{M}_{n}$ there is the
orthogonal matrix $\mathbf{T}_{n}$ and the diagonal matrix
$\mathbf{\Lambda}_{n}$ such that $\mathbf{M}_{n} =
\mathbf{T}_{n}\mathbf{\Lambda}_{n}\mathbf{T}_{n}'$. Moreover, the
diagonal matrix
$\mathbf{\Lambda}_{n}= \mathbf{T}_{n}'\mathbf{M}_{n}\mathbf{T}_{n}$
consists of the eigenvalues $\{\lambda_{i}\}$ of $\mathbf{M}_{n}$
\cite[Ch. 4.7-9]{Bellman}.

Also, for any orthogonal matrix $\mathbf{T}_{n}$, the distribution of
$\mathbf{T}_{n}'\boldsymbol{\xi}_{n}$ (for simple hypothesis
$\mathcal H_{0}$ from \eqref{mod1aa}) remains the same as of
$\boldsymbol{\xi}_{n}$. Therefore, multiplying both sides of
\eqref{mod1aa} by $\mathbf{T}_{n}'$, we reduce model \eqref{mod1aa}
to the equivalent model with the diagonal matrix $\mathbf{\Lambda}_{n}$.
In other words, model \eqref{mod1aa} takes the form
\begin{equation}\label{mod1b}
\begin{gathered}
\mathcal H_{0}: x_{i} = \xi_{i}, \qquad
\xi_{i} \sim {\mathcal N}(0,1), \\
\mathcal H_{1}: x_{i} = \eta_{i} = \lambda_{i}^{1/2}\xi_{i},
\quad i=1,\ldots,n.
\end{gathered}
\end{equation}
Then we have
\begin{equation}\label{deGaus3}
\begin{gathered}
\ln\frac{d{\mathbf P}_{n}}{d{\mathbf Q}_{n}}({\mathbf x}) =
\frac{1}{2}\sum\limits_{i=1}^{n}\left[\ln \lambda_{i} +
\left(\frac{1}{\lambda_{i}}-1\right)x_{i}^{2}\right]
\end{gathered}
\end{equation}
and
\begin{equation}\label{deGaus4}
\begin{gathered}
D(\mathbf{P}_{n}||\mathbf{Q}_{n}) =
\frac{1}{2}\sum_{i=1}^{n}\left(\ln\lambda_{i} +
\frac{1}{\lambda_{i}}-1\right).
\end{gathered}
\end{equation}
Also
\begin{equation}\label{deGaus4a}
\begin{gathered}
r_{1}^{2}({\mathbf P}_{1}^{n},{\mathbf Q}_{1}^{n}) =
{\mathbf E}_{{\mathbf P}_{n}}
\left[D({\mathbf P}_{n}||{\mathbf Q}_{n}) -
\ln\frac{d{\mathbf P}_{n}}{d{\mathbf Q}_{n}}({\mathbf x})\right]^{2} =
\\ = \frac{1}{4}\mathbf{E}\left[\sum_{i=1}^{n}
\left(\frac{1}{\lambda_{i}}-1\right)
(\xi_{i}^{2}-1)\right]^{2} = \\
= \frac{1}{4}\mathbf{E}\sum_{i=1}^{n}
\left(\frac{1}{\lambda_{i}}-1\right)^{2}
(\xi_{i}^{2}-1)^{2} = \frac{1}{2}\sum_{i=1}^{n}
\left(\frac{1}{\lambda_{i}}-1\right)^{2}.
\end{gathered}
\end{equation}
For simplicity, we assume that for a constant $C > 0$ and large $n$
the following condition is satisfied
\begin{equation}\label{deGaus4b}
\begin{gathered}
\sum_{i=1}^{n}\left(\frac{1}{\lambda_{i}}-1\right)^{2} \leq
C^{2}\sum_{i=1}^{n}\left(\ln\lambda_{i} +
\frac{1}{\lambda_{i}}-1\right).
\end{gathered}
\end{equation}
Then, by \eqref{Stein11}, \eqref{Stein14} and
\eqref{deGaus4}-\eqref{deGaus4b} we have
\begin{equation}\label{Stein11h}
\begin{gathered}
-\frac{D({\mathbf P}_{n}||{\mathbf Q}_{n})+1}{1-\alpha} \leq
\ln\beta(\alpha) \leq \\
\leq -D({\mathbf P}_{n}||{\mathbf Q}_{n}) +
C\alpha^{-1/2}\sqrt{D({\mathbf P}_{n}||{\mathbf Q}_{n})},
\end{gathered}
\end{equation}
which yields an estimate to the convergence rate in \eqref{Stein01}.

{\bf Example 3 (independent random variables)}. Consider a natural
generalization of model \eqref{mod1}, when observations
${\mathbf x}_{1}^{n} = (x_{1},\ldots,x_{n}) \in {\mathbf R}^{n}$ are
independent random variables with different distributions
$\{{\mathbf P}_{i}\}$ or $\{{\mathbf Q}_{i}\}$, $i=1,\ldots,n$:
\begin{equation}\label{example1}
\begin{gathered}
\mathcal H_{0}: {\mathbf x}_{1}^{n} \sim {\mathbf P}_{1}^{n} =
{\mathbf P}_{1}\times \ldots {\mathbf P}_{n}, \\
\mathcal H_{1}: {\mathbf x}_{n} \sim {\mathbf Q}_{1}^{n} =
{\mathbf Q}_{1}\times \ldots {\mathbf Q}_{n}.
\end{gathered}
\end{equation}
Since
\begin{equation}\label{SteinPQ1}
\begin{gathered}
D({\mathbf P}_{1}^{n}||{\mathbf Q}_{1}^{n}) =
\sum\limits_{i=1}^{n}D({\mathbf P}_{i}||{\mathbf Q}_{i}),
\end{gathered}
\end{equation}
the lower bound \eqref{Stein11} takes the form
\begin{equation}\label{Stein11aa}
\begin{gathered}
\ln\beta \geq -\frac{1}{(1-\alpha)}\left[
D({\mathbf P}_{1}^{n}||{\mathbf Q}_{1}^{n})+1\right].
\end{gathered}
\end{equation}
We also have for the remaining term
$r_{1}({\mathbf P}_{1}^{n},{\mathbf Q}_{1}^{n})$ from \eqref{rem1}
\begin{equation}\label{Stein11b}
\begin{gathered}
r_{1}^{2}({\mathbf P}_{1}^{n},{\mathbf Q}_{1}^{n}) \leq
{\mathbf E}_{{\mathbf P}_{1}^{n}}\left[
D({\mathbf P}_{1}^{n}||{\mathbf Q}_{1}^{n})
-\ln\frac{d{\mathbf P}_{1}^{n}}{d{\mathbf Q}_{1}^{n}}({\mathbf x})
\right]^{2} = \\
= \sum\limits_{i=1}^{n}
{\mathbf E}_{{\mathbf P}_{i}}\left[D({\mathbf P}_{i}||{\mathbf Q}_{i})
-\ln\frac{d{\mathbf P}_{i}}{d{\mathbf Q}_{i}}(x_{i})\right]^{2} = \\
= \sum\limits_{i=1}^{n}\left[{\mathbf E}_{{\mathbf P}_{i}}\left(
\ln\frac{d{\mathbf P}_{i}}{d{\mathbf Q}_{i}}(x_{i})\right)^{2} -
D^{2}({\mathbf P}_{i}||{\mathbf Q}_{i})\right].
\end{gathered}
\end{equation}
For simplicity, we assume that for a constant $C > 0$ and all
sufficiently large $n$ the following condition is satisfied
\begin{equation}\label{Stein11c}
\begin{gathered}
\sum\limits_{i=1}^{n}\left[{\mathbf E}_{{\mathbf P}_{i}}\left(
\ln\frac{d{\mathbf P}_{i}}{d{\mathbf Q}_{i}}(x_{i})\right)^{2} -
D^{2}({\mathbf P}_{i}||{\mathbf Q}_{i})\right] \leq \\
\leq C^{2}\sum\limits_{i=1}^{n}D({\mathbf P}_{i}||{\mathbf Q}_{i}) =
C^{2}D({\mathbf P}_{1}^{n}||{\mathbf Q}_{1}^{n}).
\end{gathered}
\end{equation}
Then, by \eqref{Stein11b}-\eqref{Stein11c}
\begin{equation}\label{Stein11d}
\begin{gathered}
{\mathbf E}_{{\mathbf P}_{1}^{n}}\left[
D({\mathbf P}_{1}^{n}||{\mathbf Q}_{1}^{n})
-\ln\frac{d{\mathbf P}_{1}^{n}}{d{\mathbf Q}_{1}^{n}}({\mathbf x})
\right]^{2} \leq C^{2}D({\mathbf P}_{1}^{n}||{\mathbf Q}_{1}^{n}).
\end{gathered}
\end{equation}
Therefore,  by \eqref{Stein11} and \eqref{Stein14} we have for $\ln\beta(\alpha)$
\begin{equation}\label{Stein11f}
\begin{gathered}
-\frac{D({\mathbf P}_{1}^{n}||{\mathbf Q}_{1}^{n})}{1-\alpha} \leq
\ln\beta(\alpha) \leq \\
\leq -D({\mathbf P}_{1}^{n}||{\mathbf Q}_{1}^{n}) +
C\alpha^{-1/2}\sqrt{D({\mathbf P}_{1}^{n}||{\mathbf Q}_{1}^{n})},
\end{gathered}
\end{equation}
which estimates the convergence rate in \eqref{Stein01}.

The author would like to thank Kutoyants Yu. A. and anonymous reviewers for
useful discussions and constructive critical remarks, which improved the paper.

\end{document}